# Impossibility of $n-1$-strong-equllibrium for Distributed Consensus with Rational Agents

Amit Jacob-Fanani

*Tel-Aviv University*

amitj@mail.tau.ac.il

Itay Harel

*Tel-Aviv University*

itayharel@mail.tau.ac.il

**Abstract**

An algorithm for $n-1$-strong-equillibrium for distributed consensus in a ring with rational agents was proposed by Afek *et al.* [1]. A proof of impossibility of $n-1$-strong-equillibrium for distributed consensus in every topology with rational agents, when $n$ is even, is presented.

Furthermore, we show that the only algorithm which can solve the problem when $n$ is odd is the one proposed in [1].

Finally, we prove that the proposed algorithm provides a $n-2$-strong-equillibrium in a synchronous ring when $n$ is even.

## I. INTRODUCTION

The problem of distributed consensus, where $n$ processors, some of which may be faulty, have to agree on the same value which is also an input to one of them, has been well studied [3-7]**.** More recently, there is growing interest in the model of distributed computing with rational agents which may deviate from a distributed algorithm in order to affect the result of the algorithm [8-12].

The main concern regarding distributed computation with rational agents is to provide an algorithm which is resilient to rational agents, meaning that rational agents would gain nothing by deviating from the protocol. In order to address this more formally, Afek *et al*. use the following definitions (refer to [1] for full details):

**Definition 1** (Utility Function). *Let A be a distributed algorithm, and O be the set of all possible final states of A. Let $p$ be an agent, then $u_p: O \to \mathbb{R}$ is the utility function, such that $\forall o \in O$, $u_p(o)$ is the profit $p$ would gain from reaching the final state $o$.*

**Definition 2** (Solution Preference). *Let A be a distributed algorithm, and O be the set of all possible final states of A. Let $p$ be an agent, and $u_p$ be its utility function, then $u_p$ satisfy the solution preference iff $\forall o_L, o_E \in O$ such that $o_L$ is a legal final state of A and $o_E$ is an erroneous final state of A, $u_p(o_L) \geq u_p(o_E)$.*

**Definition 3** (K-Strong-Equilibrium). *Let A be a distributed algorithm, and P the set of processors, then A reaches k-strong equilibrium iff $\forall C \subseteq P$, such that $|C| \leq k$, for every $p \in C$ the following holds:*

$$\mathbb{E}[u_p | C \text{ deviates from } A] \leq \mathbb{E}[u_p | C \text{ runs } A]$$

The problem of distributed consensus with rational agents is addressed in [1] and in [2]. In the model proposed in [1], some processors may prefer a certain outcome of the protocol, i.e. that all processors eventually agree on the value 1 or that all agree on the value 0.

Afek *et al.* propose in [1] an algorithm providing a $n-1$-strong-equilibrium for the problem of distributed consensus with rational agents in a synchronous ring. The algorithm is presented in IV.

In this paper we show that this algorithm fails in case $n$ is even. Moreover, we show that a $n-1$-strong-equilibrium is unattainable when $n$ is even, in every topology (II). Nonetheless, we prove that it is the only algorithm that provides a $n-1$-strong-equilibrium in a synchronous ring, when $n$ is odd (III), and that the algorithm does provide a $n-2$-strong-equilibrium in a synchronous ring (IV).

## II. IMPOSSIBILITY OF $n-1$-STRONG-EQUILIBRIUM WHEN $n$ IS EVEN

**Definition 4** (Input Cheaters). *Let A be an algorithm for distributed consensus and let C be a coalition of cheaters, then C are input cheaters iff C run A but may act as if they have a different set of inputs than they are actually given.*

**Theorem 1** (Impossibility of $n-1$-Strong-Equilibrium). *Assuming uniform distribution over the inputs, there is no algorithm for distributed consensus, in every topology, providing a $n-1$-strong-equilibrium when $n$ (the number of nodes) is even.*

*Proof.* Assume by contradiction that there is an algorithm $A$ that provides a $n-1$-strong-equillibrium.

Given a coalition of size $n-1$, denote the single node which is not part of this coalition by $v$, and the group of nodes in the coalition (i.e. $V/\{v\}$) by $C$.

We will show that even when we bound $C$ to be input cheaters, $A$ will fail to provide a $n-1$-strong-equillibrium. We will use the following notations:

1. $D(u)$ – The value decided by a node $u$ when it finishes to run $A$.
2. $J(u)$ – The input that a node $u \in C$ pretends to have.
3. $I(u)$ – The input a node $u$ is given (before it starts running $A$).
4. $I(C) = \big(I(u)\big)_C$, i.e. the set of inputs given to the nodes of $C$.
5. $J(C) = \big(J(u)\big)_C$, i.e. the set of inputs that the nodes of $C$ pretend to have.
6. For $j: C \to \{0,1\}$ define $ones(j) = |\{u \in C | j(u) = 1\}|$

Assume $I(v) = 1$ and $\forall u \in C: J(u) = 1$, then by *validity* attribute of consensus $D(v) = 1$ is deduced (since $v$ cares only about $J(C)$ and cannot know that $I(C) \neq J(C)$).

Formally,

$$(1) \quad \mathbb{P}\big(D(v) = 1 \big| I(v) = 1, ones(J(C)) = |C|\big) = 1$$

And similarly,

$$(2) \quad \mathbb{P}\big(D(v) = 0 \big| I(v) = 0, ones(J(C)) = 0\big) = 1$$

Define $\alpha(j)$ the parity of $ones(j)$, i.e. $\alpha(j) \stackrel{\text{def}}{=} ones(j) \bmod 2$.

**Lemma 1.1.** *Assume C runs A with $J(C) = j$, then $D(v) = \neg\big(I(v) \oplus \alpha(j)\big)$.*

*Proof.* We will prove the lemma by backwards induction on $ones(j)$. If $ones(j) = n-1$, then $\alpha(j) = 1$.
If $I(v) = 1$ then using (1) we conclude $D(v) = 1$, and indeed $1 = \neg(1 \oplus 1)$.
Otherwise, $I(v) = 0$. Assume that $I(C) = (0, \dots, 0)$ and that $\forall p \in C: u_p(0) = 0, u_p(1) = 1$ i.e. every player in $C$ prefers that $D(v) = 1$.

If $C$ did not cheat, their expected utilization was:

$$(3) \quad \mathbb{E}\big[u_p \big| J(C) = I(C)\big] = \mathbb{P}(D(v) = 1 | J(C) = (0)_C) =_1 \mathbb{P}(I(v) = 1) \cdot \mathbb{P}(D(v) = 1 | J(C) = (0)_C, I(v) = 1) \leq \mathbb{P}(I(v) = 1) = \frac{1}{2}$$

Where $=_1$ is a result of (2).

On the other hand, if $C$ cheated with $ones(j) = n-1$, i.e. $j = (1)_C$, then their expected utilization was:

$$\begin{aligned}
(4) \quad \mathbb{E}\big[u_p \big| J(C) = (1)_C\big] &= \mathbb{P}(D(v) = 1 | J(C) = (1)_C) = \mathbb{P}(D(v) = 1 | J(C) = (1)_C, I(v) = 1) \cdot \\
&\quad \mathbb{P}(I(v) = 1) + \mathbb{P}(D(v) = 1 | J(C) = (1)_C, I(v) = 0) \cdot \mathbb{P}(I(v) = 0) = \tfrac{1}{2} + \\
&\quad \tfrac{1}{2}\mathbb{P}(D(v) = 1 | J(C) = (1)_C, I(v) = 0)
\end{aligned}$$

Since $A$ provides a $n - 1$-strong-equillibrium then $C$ should not benefit from cheating. Formally,

$$(5) \quad \forall p \in C \colon \mathbb{E}\big[u_p \big| I(C) = J(C)\big] \geq \mathbb{E}\big[u_p \big| J(C) = (1)_C\big]$$

Combining (3), (4) and (5) we have

$$(6) \quad \forall p \in C \colon \tfrac{1}{2} \geq \mathbb{E}\big[u_p \big| I(C) = J(C)\big] \geq \mathbb{E}\big[u_p \big| J(C) = (1)_C\big] = \tfrac{1}{2} + \tfrac{1}{2}\mathbb{P}(D(v) = 1 | J(C) = (1)_C, I(v) = 0)$$

Hence, $\mathbb{P}(D(v) = 1 | J(C) = (1)_C, I(v) = 0) = 0$, thus $D(v) = 0$. Indeed $0 = \neg(0 \oplus 1)$.

We will assume the statements hold for $ones(j) = k$ and prove it for $ones(j) = k - 1$.

Consider the first case $\alpha(j) = 0$, i.e. $k \bmod 2 = 1$.

If $I(v) = 1$ then consider a node $u \in C$ such that $J(u) = 0$ (there must be such $u$ because $ones(j) < n - 1$). Note that since $C$ are input cheaters, $u$ is in the same situation as $v$ is in case $I(v) = 0$ and $ones(j) = k$, except that $u$ is in the coalition and $v$ is not. Of course, $v$ cannot know this, therefore applying the induction hypothesis where $ones(j) = k$ and $I(v) = 0$, $v$ may locally deduce that $D(u) = 0$. Using *agreement* property of consensus, we have $D(v) = 0$. Indeed $0 = \neg(1 \oplus 0)$.

If $I(v) = 0$ then we can assume that $I(C) = (1)_C$ and that $\forall p \in C \colon u_p(0) = 1, u_p(1) = 0$ i.e. every player in $C$ prefers that $D(v) = 0$. If $C$ did not cheat, their expected utilization was:

$$\mathbb{E}\big[u_p \big| J(C) = I(C)\big] = \mathbb{P}(D(v) = 0 | J(C) = (1)_C) =_1 \mathbb{P}(D(v) = 0 | J(C) = (1)_C, I(v) = 0) \cdot \mathbb{P}(I(v) = 0)$$
$$\leq \mathbb{P}(I(v) = 0) = \frac{1}{2}$$

Where $=_1$ is a result of (1).

Notice that from the case $I(v) = 1$ above, $\mathbb{P}(D(v) = 0 | J(C) = j, I(v) = 1) = 1$ follows. Therefore, if $C$ did cheat, the expected utilization of $C$ was:

$$\begin{aligned}
\mathbb{E}\big[u_p \big| J(C) = j\big] &= \mathbb{P}(D(v) = 0 | J(C) = j) \\
&= \mathbb{P}(D(v) = 0 | J(C) = j, I(v) = 1) \cdot \mathbb{P}(I(v) = 1) + \mathbb{P}(D(v) = 0 | J(C) = j, I(v) = 0) \\
&\quad \cdot \mathbb{P}(I(v) = 0) = \tfrac{1}{2} + \tfrac{1}{2}\mathbb{P}(D(v) = 0 | J(C) = j, I(v) = 0)
\end{aligned}$$

Since $A$ provides a $n - 1$-strong-equillibrium then $C$ should not benefit from cheating. Formally,

$$\forall p \in C \colon \tfrac{1}{2} \geq \mathbb{E}\big[u_p \big| I(C) = J(C)\big] \geq \mathbb{E}\big[u_p \big| J(C) = j\big] = \tfrac{1}{2} + \tfrac{1}{2}\mathbb{P}(D(v) = 0 | J(C) = j, I(v) = 0)$$

Thus, we have $\mathbb{P}(D(v) = 0 | J(C) = j, I(v) = 0) = 0$, so $D(v) = 1$. Indeed $1 = \neg(0 \oplus 0)$.

The second case $\alpha(j) = 1$, is symmetric.

∎

In case $ones(j) = 0$ and $I(v) = 0$, by Lemma 1.1 we have that $D(v) = \neg(0 \oplus 0) = 1$, contradicting (2).

∎

### III. SINGULARITY OF THE SOLUTION ALGORITHM WHEN $n$ IS ODD

Assume $A$ is an algorithm providing a $n - 1$-strong-equilibrium when $n$ is odd. We can repeat the proof of Lemma 1.1, replacing the claim with $D(v) = I(v) \oplus \alpha(j)$.

The proof works exactly the same, but the result of this lemma doesn't contradict (2).

Instead, we have that even number of nodes with an input equal to 1 results in $D(v) = 0$ and odd number of nodes with an input equal to 1 results in $D(v) = 1$. Thus, if we look at $A$ as a function that maps between the nodes' inputs and the output of the distributed consensus, $A$ is exactly equal to the algorithm given in [1].

## IV. A $n - 2$-STRONG-EQULLIBRIUM

**Theorem 2** (A $n - 2$-Strong-Equilibrium Algorithm) *Assuming uniform distribution over the inputs and a synchronous ring of even size, the algorithm proposed in [1] works and provides a $n - 2$-strong-equilibrium.*

*Proof.* To simplify, we shall start by assuming the ring is unidirectional. A method to discard this assumption regardless of the cheating coalition is shown in [2].

We shall present the algorithm proposed in [1]:

**Algorithm 1** (Afek *et al.* [1]):

```
         do_consensus(input, id):
1                my_rand := rand()
2                ids_array.append(<id, input>)
3                rand_sum := my_rand
4                input_sum := input
5                Send(<input, id, my_rand>, out_interface)
6                for i = 1,…,n-1:
7                        tmp := Recv(in_interface)
8                        if tmp = null or not (tmp.input ∈ {0, 1}):
9                                REPORT CHEATER
10                       ids_array.append(<tmp.id, tmp.input>)
11                       rand_sum += tmp.rand
12                       input_sum += tmp.input
13                       Send(tmp, out_interface)
14               end for
15               tmp := Recv(in_interface)
16               if tmp.input ≠ input or tmp.id ≠ id or tmp.rand ≠ my_rand:
17                       REPORT CHEATER
18               sort ids_array by id
19               if ∃id ∈ ids_array with ids_array.count(id) ≠ 1:
20                       REPORT CHEATER
21               leader := ids_array[rand_sum % n]
22               input_sum += leader.input
23               if input_sum % 2 == 1:
24                       return 1
25               else:
26                       return 0
         end do_consensus
```

**Lemma 2.1** *If all nodes in the ring comply with the algorithm, it maintains both validity and agreement, and $\mathbb{P}(ring\ decides\ 0) = \mathbb{P}(ring\ decides\ 1) = \frac{1}{2}$.*

*Proof.* Addressing validity – if every node got 0 as an input, then for every node in the ring $u.input\_sum$ will be 0, meaning that $u.input\_sum\ mod\ 2 = 0$ resulting in all nodes agreeing on 0. If every node got 1 as input, then for every node in the ring $u.input\_sum$ will be $n + 1$, and since $n$ is even – it will result in 1 as consensus. If some nodes got 0 and others got 1, every outcome is acceptable in terms of validity.

Addressing agreement – if all nodes comply then every node after line 18 will have the exact same values for $rand\_sum$, $input\_sum$ and $ids\_array$ (follows from a simple induction), which will result in all nodes agreeing on the same value.

Addressing $\mathbb{P}(ring\ decides\ 0) = \mathbb{P}(ring\ decides\ 1) = \frac{1}{2}$ - we shall prove the following claims by induction on the number of nodes:

1. Given nodes $v_1, v_2, \ldots v_{2k}$: $\mathbb{P}(\sum_{i=1}^{2k} v_i.input\ mod\ 2 = 0) = \mathbb{P}(\sum_{i=1}^{2k} v_i.input\ mod\ 2 = 1) = \frac{1}{2}$
2. Given nodes $v_1, v_2, \ldots v_{2k}$;
   $\mathbb{P}(\sum_{i=1}^{2k} v_i.input + leader.input\ mod\ 2 = 0) = \mathbb{P}(\sum_{i=1}^{2k} v_i.input + leader.input = 1) = \frac{1}{2}$

If $n = 2$ (we shall denote the nodes as $v_1, v_2$): Addressing the first claim – there are 4 possible inputs that can be given to the nodes - <0,0>, <0,1>, <1,0>, <1,1>. In half of them the sum of all inputs is even, in the other half – odd.

Addressing the second claim - if $v_1$ and $v_2$ both got 0 as an input – the sum (including the leader) is even. If both got 1 – the sum (including the leader) is odd. If $v_1$ got 0 and $v_2$ got 1 – then if $v_1$ is elected leader they will decide on 1, else they will decide on 0. The exact opposite result will occur if $v_1$ got 1 and $v_2$ got 0. In conclusion – in exactly half of the cases the sum (including leader) is even, and in the other half – odd.

Now, assume the above claims applies to all rings of size $n = 2(k-1)$. We shall prove it applies to a ring of size $n = 2k$.

Addressing claim 1 – By applying the same analysis done in the base case for nodes $n_{2k}, n_{2k-1}$ and by using the induction hypothesis, the claim is proven.

Addressing claim 2 – Denote the nodes as $v_1, v_2, \ldots v_{2k}$.

If $v_{2k}$ and $v_{2k-1}$ aren't elected as leader, then their contribution to $input\_sum$ is odd (exactly 1) half of the times and even (0 or 2) the other half. By induction hypothesis (claim 2), we know that if one of the nodes $v_1, v_2, \ldots v_{2k-2}$ was elected leader, then in half of the cases their sum of inputs (including leader) is 0, and in the other half – it is 1. Thus the contribution of $v_{2k-1}, v_{2k}$ maintains the probability of 0 or 1 to be decided.

If $v_{2k}$ or $v_{2k-1}$ are elected as leader, then by applying claim 1 on $v_1, v_2, \ldots v_{2k-2}$ and claim 2 on $v_{2k-1}, v_{2k}$, the claim is proven.

From claim 2, it is clear that $\mathbb{P}(ring\ decides\ 0) = \mathbb{P}(ring\ decides\ 1) = \frac{1}{2}$.

∎

Each node in the ring is initialized with 3 values - $<id, input, random>$. Denote $Trip(u)$ - the triplet of values with which $u$ is initialized in the beginning of the algorithm; and $Trip(v, u)$ – a triplet of values received by $u$ during the algorithm in which $Trip(v, u).id = v$. Note that $input$ and $random$ may or may not be the actual values with which $v$ was initialized in the beginning. Furthermore, the id $v$ may or may not be an id of an actual node in the ring.

Let $v \in C, u \notin C$ such that $v$ is the closest member of the coalition to $u$ from $u$'s $in\_interface$. Also, let $m$ be the number of nodes between $v$ and $u$, including u. It is clear that $v$ (as a member of the coalition) may lie regarding $n - m$ of the triplets sent to $u$ (thus, giving the coalition control over $n - m$ messages) but cannot affect $m$ of the triplets received by $u$ ($Trip(u)$ and the triplets of the $m - 1$ nodes between $v$ and $u$). Further, $v$ receives these $m$ triplets after it had sent all $n - m$ triplets controlled by him. Of course, $v$ cannot change the triplets it sends after it had sent them, which implies that any information that $u$ received from $v$ is independent of any information that $v$ received from $u$. In order to reflect this conclusion, we define the following:

1. $R(v, u)$ – the set of nodes between $v$ and $u$, including $u$.
2. $Lie(u) = \{Trip(w, u) | w \notin R(v, u)\}$. These triplets are fully controlled by $C$.
3. $Truth(u) = \{Trip(w, u) | w \in R(v, u)\}$. These aren't controlled by C.
4. $InputT(u) = \sum_{Trip\ \in Truth(u)} Trip.input$
5. $InputL(u) = \sum_{Trip\ \in Lie(u)} Trip.input$
6. $RandomT(u) = \sum_{Trip\ \in Truth(u)} Trip.random$
7. $RandomL(u) = \sum_{Trip\ \in Lie(u)} Trip.random$

**Lemma 2.2** *Given a coalition of cheaters denoted $C$, and a node $u \notin C$, then every id in $u.ids\_array$ has the same probability of $\frac{1}{n}$ to be elected as leader by $u$, regardless of the actions taken by $C$.*

*Proof.* This lemma has been explored and proved in [1], [2]. In short – $Trip(u) \in Truth(u) \neq \phi$ which means that $RandomT(u)$ is uniformly distributed. Since $u.rand\_sum$ includes $RandomT(u)$ in the sum, $u.rand\_sum \bmod n$ is uniformly distributed as well, which means that every value between 0 to $n-1$ has an equal chance to be selected as an index to $u.ids\_array$.
∎

**Lemma 2.3** *Given a coalition of cheaters denoted $C$, and a node $u \notin C$, if the leader elected by $u$ is $p \neq u$ then $P(u.consensus = 1) = P(u.consensus = 0) = \frac{1}{2}$.*

*Proof.* From the algorithm, after line 22 we have:
$$u.input\_sum = InputT(u) + InputL(u) + Trip(u.leader, u).input$$
Since we have assumed a uniform distribution over the inputs, the probability that $inputT(u)$ is odd is equal to the probability that it is even, i.e. $\frac{1}{2}$. This implies that $input\_sum$ has the same probability to be odd and even, regardless of the parity of $InputL(u) + Trip(u.leader, u)$. This proves Lemma 2.3.
∎

**Lemma 2.4** *Assume a coalition of cheaters denoted $C$, and a node $u \notin C$, where $u.input\_interface \in C$. If the leader elected by $u$ is $u$, then $C$ has full control over the decision of $u$.*

*Proof.* Since $u.input\_interface$ is a cheater, $InputT(u) = u.input$. After line 22:
$$u.input\_sum = u.input + InputL(u) + Trip(u.leader, u).input = 2 \cdot u.input + InputL(u)$$
Hence the result of the consensus algorithm is:
$$u.consensus = u.input\_sum \bmod 2 = (2 \cdot u.input + InputL(u)) \bmod 2 = InputL(u) \bmod 2$$
Which means that $u$'s decision depends only on the values sent to $u$ from $u.input\_interface$.
∎

**Lemma 2.5** *Assume a coalition of cheaters denoted $C$, with $|C| = n - 2$. Also, assume $u_1, u_2 \notin C$ aren't neighbors in the ring. In this case, $C$ has no incentive to cheat.*

*Proof.* Assume w.l.o.g, that $C$ prefers the consensus to be 1. For convenience, we shall divide our analysis to the following three cases:

**Case 1**: $u_1, u_2$ received different $random$ or $id$ values during the course of the algorithm. Then we can analyze this case as if $u_1$ and $u_2$ were on different rings, thus their decisions are independent.

By 2.2:

$$(1) \quad \mathbb{P}(u_1.leader = u_1) = \frac{1}{n}, \mathbb{P}(u_1.leader \neq u_1) = \frac{n-1}{n}$$

It follows from 2.3 and 2.4 that:

$$(2) \quad \mathbb{P}(u_1.consensus = 1 \mid u_1.leader \neq u_1) = \frac{1}{2}$$

$$(3) \quad \mathbb{P}(u_1.consensus = 1 \mid u_1.leader = u_1) = 1$$

Combining (1), (2) and (3):

(4) $\mathbb{P}(u_1.consensus = 1) = \frac{1}{n} + \frac{n-1}{2n}$

The exact same result applies to $u_2$. As we explained, in this case $u_1.consensus$ is independent of $u_2.consesnsus$, therefore the probability for a consensus to be reached over 1 is:

(5) $\mathbb{P}(u_1.consensus = 1 \wedge u_2.consensus = 1) = \left(\frac{1}{n} + \frac{n-1}{2n}\right)^2 = \frac{1}{4}\left(1 + \frac{1}{n}\right)^2 \leq_1 \frac{1}{4}\left(1 + \frac{1}{4}\right)^2 < \frac{1}{2}$

Where $\leq_1$ is a result of $n \geq 4$ (there are 2 non cheaters and an even non-zero number of coalition members). Therefore $C$ has no incentive to cheat in Case 1.

**Case 2:** The $random$ and $id$ values received by $u_1$ and $u_2$ during the course of the algorithm are identical, but the $input$ values are different. In this case, the leader elected by $u_1$ is the same as the one elected by $u_2$. For convenience, we shall denote it as $leader$, since both nodes choose the same leader in this case.

By Lemma 2.2:

(6) $\mathbb{P}(leader \neq u_1, u_2) = \frac{n-2}{n}$, $\mathbb{P}(leader = u_1) = \frac{1}{n}$, $\mathbb{P}(leader = u_2) = \frac{1}{n}$

From Lemmas 2.3 and 2.4 we have:

(7) $\mathbb{P}(u_1.consensus = u_2.consensus = 1 \mid leader \neq u_1, u_2) = \frac{1}{4}$

(8) $\mathbb{P}(u_1.consensus = u_2.consensus = 1 \mid leader = u_1) = \frac{1}{2}$

(9) $\mathbb{P}(u_1.consensus = u_2.consensus = 1 \mid leader = u_2) = \frac{1}{2}$

By combining (6) – (9), we can derive:

(10) $\mathbb{P}(u_1.consensus = 1 \text{ and } u_2.consensus = 1) = \frac{1}{4} * \frac{n-2}{n} + \frac{1}{2} * \frac{1}{n} + \frac{1}{2} * \frac{1}{n} = \frac{1}{4} + \frac{1}{2n} \leq \frac{1}{4} + \frac{1}{8} < \frac{1}{2}$

Thus $C$ has no incentive to cheat in Case 2 as well.

**Case 3:** Nodes $u_1$ and $u_2$ receive the exact same triplets during the course of the algorithm.

In this case at least one of the nodes will not choose itself as the leader. By Lemma 2.3 the probability of this node deciding on 1 is $\frac{1}{2}$, and this bounds the probability of a consensus over this value. Hence, $C$ has no incentive to cheat in Case 3.

By cases 1 – 3, Lemma 2.5 is proved.

∎

**Lemma 2.6** Assume a coalition of cheaters denoted $C$, and two nodes $u_1, u_2 \notin C$ such that $u_1 = u_2.input\_interface$. Then $\mathbb{P}(u_2.consensus = 1) = \mathbb{P}(u_2.consensus = 0) = \frac{1}{2}$.

*Proof.* By Lemma 2.3, if $u_2$ chooses a leader other than itself, then:

$$\mathbb{P}(u_2.consensus = 1) = \mathbb{P}(u_2.consensus = 0) = \frac{1}{2}$$

If $u_2$ chooses itself as a leader, then $u_2.input\_sum = 2 \cdot u_2.input + u_1.input + InputL(u)$

Since $\mathbb{P}(u_1.input = 1) = \mathbb{P}(u_2.input = 1) = \frac{1}{2}$ from the exact same arguments of Lemma 2.3 we have $\mathbb{P}(u_2.consensus = 1) = \mathbb{P}(u_2.consensus = 0) = \frac{1}{2}$.

∎

**Lemma 2.7** *Assume a coalition of cheaters denoted $C$ with $|C| = n - 2$, and two nodes $u_1, u_2 \notin C$ such that $u_1 = u_2.input\_interface$. Then there is no incentive for $C$ to cheat.*

*Proof.* Follows from Lemma 2.6.

∎

**Proof of Theorem 2** (A $n - 2$-Strong-Equilibrium Algorithm)
*Proof.* From Lemma 2.1 we know that the algorithm maintains both validity and agreement when every node complies, and also that $\mathbb{P}(ring\ decides\ 0) = \mathbb{P}(ring\ decides\ 1) = \frac{1}{2}$. From Lemmas 2.5 and 2.7 we know that regardless of the position of the two non-cheaters (neighbors or not) there is no incentive for cheaters to cheat. Thus, the algorithm provides a $n - 2$-strong-equilibrium.

∎

## V. CONCLUSION

In this paper, we have proved that problem of distributed consensus with a coalition of $n - 1$ rational agents cannot be solved when $n$ is even, regardless of the network topology. Furthermore, we have proved that when $n$ is odd, there can be only one method to reach consensus – the parity of the sum of all inputs.

Also, we have proved that the algorithm from [1] solves the problem of distributed consensus in a synchronous ring with a coalition of $n - 2$ rational agents, when $n$ is even. Thus, we gave both an upper and a lower bound on the maximal size of a coalition supported by a correct protocol for the problem of distributed consensus with rational agents in this topology.

## VI. ACKNOWLEDGEMENT

We would like to express our gratitude to Prof. Yehuda Afek, for his support, patience, and encouragement throughout our time as his students. We also thank Rhea Chowers for his comments on an earlier version of this paper.